# ELECTROSTATICALLY - DRIVEN RESONATOR ON SOI WITH IMPROVED TEMPERATURE STABILITY

*Archit Giridhar[1,2], Fabrice Verjus[1], Frédéric Marty[2], Alain Bosseboeuf[3], Tarik Bourouina[2]*

[1] PHILIPS Semiconductors, 2 Rue de la Girafe BP 5120 14079 Caen Cedex 5, France
[2] ESIEE / ESYCOM Lab. Cité Descartes 2 Bd Blaise Pascal 93162 Noisy-le-Grand Cedex France
[3] IEF UMR8622 CNRS – Université Paris Sud. Bât. 220 91405 Orsay Cedex France

## ABSTRACT

*This paper deals with a single-crystal-silicon (SCS) MEMS resonator with improved temperature stability. While simulations have shown that the temperature coefficient of resonant frequency can be down to 1 ppm/°C, preliminary measurements on non-optimised structures gave evidence of a temperature coefficient of 29 ppm/°C. Design, optimisation, experimental results with post process simulation and prospective work are presented.*

## 1. INTRODUCTION

The reference oscillator is among the most difficult to realize, since its stability specifications are quite demanding. The resonators for digital communication are currently built with the quartz crystal. The high quality factors and high temperature stability of such components make them more reliable in application. However quartz resonators have limited integration level and therefore their use in circuits and systems is still expensive.

On the other hand, the demonstrated MEMS resonators are theoretically sufficient to satisfy the short-term stability needs of reference oscillators, but their aging stability and their temperature stability still remains a challenge. It has been shown that the long-term aging rate of these resonators is equivalent to commercial quartz crystal resonators. Also, unlike quartz resonators, no initial aging or stabilization and no measurable hysteresis are found [1]. Indeed, the temperature excursion for an uncompensated clamped-clamped beam polysilicon micro mechanical resonator is more than 15 times higher than that of the worst AT-cut quartz crystal. Though MEMS resonators satisfy the requirements of IC-compatible tanks for use in the low phase noise oscillators and highly selective filters of communications systems [2] yet they fall short of the marks required by oscillators with respect to temperature stability. In order to achieve more stable MEMS resonators, the design technique challenges between the thermal dependence of geometrically tailored stresses and temperature variations of Young's modulus and density [3]. Further demonstrates that it is possible to reduce the temperature coefficient of the resonance frequencies by self-compensation, without any additional power consumption.

## 2. WORKING PRINCIPLE

This operation principle of the self-compensated resonator was described in a report from N'Guyen's group and demonstrated experimentally on surface micro machined polysilicon resonators.

The self-compensation of the resonance frequency method is based on reducing the effect of material properties thermal variation by the stress induced in a compensating structure. The compensating structure is designed so that it will expand faster with increasing temperature, generating a net mechanical tensile stress in the resonator beam. This tensile stress serves to increase the beam's resonance frequency and thus oppose frequency decrease caused by Young's modulus temperature dependence, resulting in a smaller (and even nil) overall frequency excursion over a given temperature range.

## 3. RESONATOR ARCHITECTURE

The flexural mode resonator beam is attached to the wings containing the comb fingers with the help of the wing support on either side (Fig. 3.1). The resonator beam is connected to a horizontal truss beam at the top end and the bottom end is anchored to the substrate. The truss beam is fixed to the compensating beams on either side of the horizontal truss beam. The other ends of the compensating beam are anchored similar to that of the resonator beam. To insure that only the resonator beam vibrated when excitation signals are applied, the outer support beams are made much wider than the resonator beam, making them rigid against lateral motions. The comb drive (not shown in Fig.3.1) employed for the electrostatic actuation of the resonator is placed in between the compensating beam and the wing with the movable fingers.

*Archit Giridhar[1,2] , Fabrice Verjus[1], Frédéric Marty[2], Alain Bosseboeuf[3], Tarik Bourouina[2]*
*Electrostatically - driven Resonator on SOI with Improved Temperature Stability*

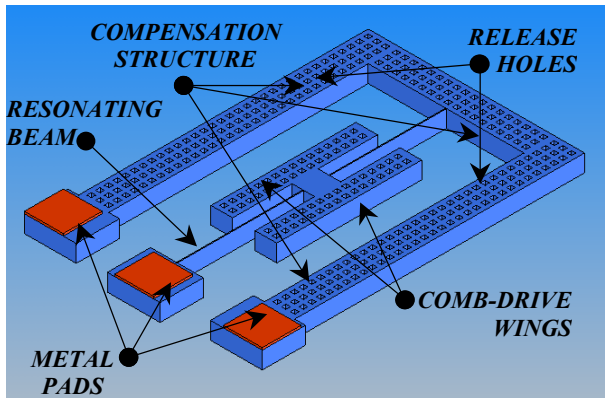

Fig. 3.1: A 3D view of the resonator

In addition to reduced temperature sensitivity, this design also features a support on the substrate that effectively reduces energy radiation into the substrate during resonance vibration [4]. Although one end of the resonator beam is rigidly anchored to the substrate, the other end is held flexible by the supporting system resulting in an end condition that is more decoupled from the substrate than a rigidly anchored end condition. This end condition, together with the use of a lateral vibration mode, effectively reduces anchor losses to the substrate, providing higher resonator Q than conventional clamped-clamped beam resonator.

### 4. SIMULATION

A simple structure of the proposed design was simulated using ANSYS © commercial software. The structure taken for simulation is shown in (fig.3.1). The simulated results give information about the rank of modes of the useful resonance, frequencies and effect of thermal variation. The simulated structure has the following characteristics given in table 1.

| | | | |
|---|---|---|---|
| Density (kg/m3) | 2330 | | |
| Young's modulus (N/m2) | 1.61e11 | | |
| Temperature coefficient of the Young's modulus (ppm/°C) | -7.406 | | |
| Thermal expansion coefficient (ppm/°C) | 2.4 | | |
| Structure thickness t (µm) | 20 | | |
| Resonator beam length L (µm) | 560 | | |
| Resonator width w (µm) | 2 | | |
| Compensating width (perforated) W (µm) | 40 | | |
| Compensating Length ratio L1/L2 | 530/560 | 540/560 | 550/560 |

Table.1: Dimensions and material properties
where $L_1$ is the length of the compensating beams and $L_2$ is the length of the resonating beam.

The simulation focused on the resonant frequency and then to optimise for the temperature stability. We can observe the lateral in-plane displacement of the resonating structure, with a slight out-of-plane flexural displacement of the compensation structure (second resonance mode). Modes 3, 4 and 5 correspond to twisting modes but were strongly rejected into the considered frequency range. The simulated result of 33.6 kHz resonance frequency was achieved with a 30 fingers equivalent geometry. The mechanically stressed zones can be observed with an energy analysis. An important stress exists at the lower resonating beam anchor and on the comb-drives support. However, the compensation structure anchor was low-stressed.

### 5. FABRICATION

#### 5.1. SOI Technology

The SOI technology consists of a two-mask process. The first one was used to pattern gold electrodes. The second one defines the geometry of the resonator mechanical structure, which was obtained by deep etching of silicon using DRIE with photo resist as a mask material. Finally the release of the structures was performed using hydrofluoric acid (HF). SOI technology provides electrical isolation between the bodies of individual SCS resonators which can be used for instance in an array implementation (for filter synthesis for instance). In general, it enables sub-micron precision fabrication of stiff SCS resonators with height to width ratio > 10 for better electro-mechanical coupling and higher frequency operation.

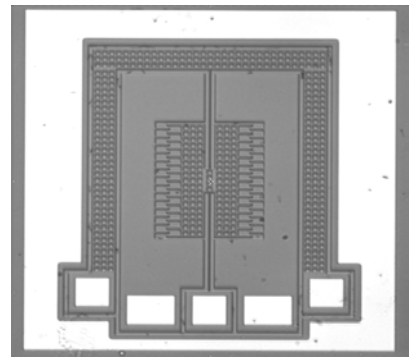

Fig. 5.1 The top view of the fabricated structure (resonator)

The SEM picture of the fabricated resonator corresponds to the structure considered for the simulation. The release etch holes were taken into consideration for the laying out the structure during fabrication of the resonator (fig.5.1).

*Archit Giridhar[1,2], Fabrice Verjus[1], Frédéric Marty[2], Alain Bosseboeuf[3], Tarik Bourouina[2]*
*Electrostatically - driven Resonator on SOI with Improved Temperature Stability*

## 6. EXPERIMENTAL RESULTS

The experimental data shows that the resonator was quite stable to a large thermal variation for a range of 25°C to 80°C. Measurements were performed by an optical vibrometer using the knife edge method [5], in order to have access to in-plane vibrations. Measurements were performed in vacuum ($7 \times 10^{-2}$ mbar) in order to emphasize the quality factor. The results were plotted by simultaneously mapping the vibration [6]. The measured resonant frequency was 6320 Hz at room temperature (25°C). When the temperature was varied up to 80°C the resonant frequency shifted to 6330 Hz. This slight change by 10 Hz in the resonant frequency was equivalent to 29 ppm/°C, which was in the same order of magnitude of the simulation results on non-optimised structures.

A study on the experimental results has been proposed to analyse the deviation in the theoretical and experimental results. Measurements are now underway on devices having different truss and compensating beam dimensions and also different overall dimensions of the resonator. As already seen in the design and simulation sections, the ratio of the length of the compensating beam and the resonating beam were expected to have an impact on the temperature coefficient of the resonance frequency.

## 7. POST PROCESS SIMULATION

In order to analyse the results of the experiment, i.e. measurements made on the fabricated resonator the post process simulation of the structure was done. During this simulation all parameters were considered including the release etch holes. The structure identified in the simulation was similar to the structure fabricated. This simulation was done to study the deviations in the experimental results in contrast with the previously simulated results. By this simulation procedure, the idea of achieving the resonator with improved temperature stability at the required frequency (32.786 KHz) is enhanced.

The various parameters were considered for the post process simulation. The dimensions of the structure (fabricated resonator) observed with the interferometer were defined for simulation along with the etch holes imposed on the resonator for the release of the structures. The effect of variation in resonant frequency with temperature variation from (-40 to 80) i.e. Mechanical simulations between –40°C to 80°C

The first level of post process simulation concentrates more on the study to analyse the difference in the resonance frequency from 33.6 KHz to 6.3 KHz. at 25°C. Later, the second level of simulation was to relate the experimental results and the simulated results based on the real dimensions of the fabricated structure obtained from the SEM pictures. Finally, third level of simulation focused on the stability on resonant frequency with a variation in temperature.

### 7.1. Effect of release etch holes

In the first level of the post process simulation a few etch holes were introduced in the structure (fig.7.1.1). The release etch holes were consideration to evaluate the effect on the structure i.e. the variation in the resonant frequency and the temperature stability of the structure.

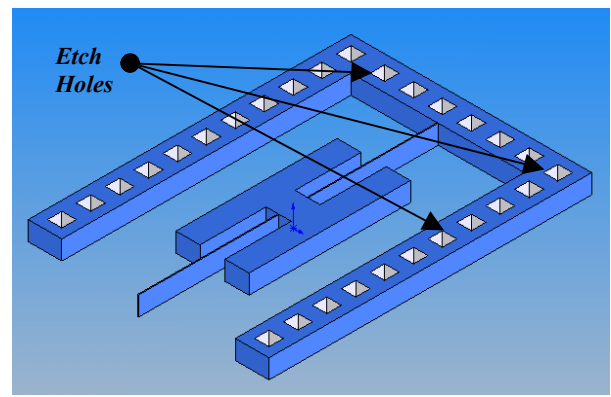

*Fig. 7.1.1: A view of the simulated structure*

The simulations have shown that the required flexural resonant mode (Fig.7.1.2) was achieved typically at a frequency of 8335 Hz. This mode was at the first rank, whose frequency was far from higher order modes: more than 3700 Hz frequency difference with the second mode. This shows that the effect of the etch holes reduced the stiffness of the structure resulting in the lowering of the resonant frequency. The curves shown in (Graph. 7.1.1) summarize the simulation results for the temperature dependence of the first flexural resonance frequency. The relative variation of frequency versus temperature has either a positive or negative slope depending on the value of the beam length ratio L1/L2. One can also see that the minimum slope was obtained for the ratio L1/L2 = 540/560 (=0.964), corresponding to a theoretical frequency temperature coefficient of –1.6 ppm/°C for this optimised structure. Other values for the frequency temperature coefficients of non-optimised structures were –10.9 ppm/°C for L1/L2 = 530/560 (=0.946) and 8.2 ppm/°C for L1/L2 = 550/560 (=0.982).

*Archit Giridhar[1,2], Fabrice Verjus[1], Frédéric Marty[2], Alain Bosseboeuf[3], Tarik Bourouina[2]*
*Electrostatically - driven Resonator on SOI with Improved Temperature Stability*

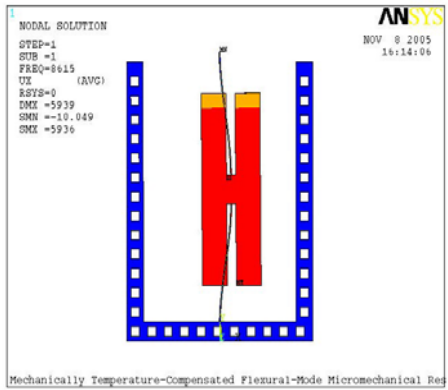

*Fig. 7.1.2:  The simulation results on the structure*

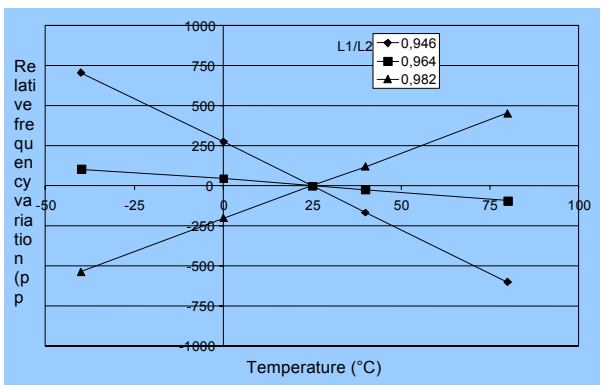

*Graph. 7.1.1: The variation of temperature stability with respect to the variation in ratio (L₂/L₁)*

## 7.2. Deviation Analysis

The second level of simulation was done on the structure approximately identical to that of the fabricated structure (fig.7.2.1).

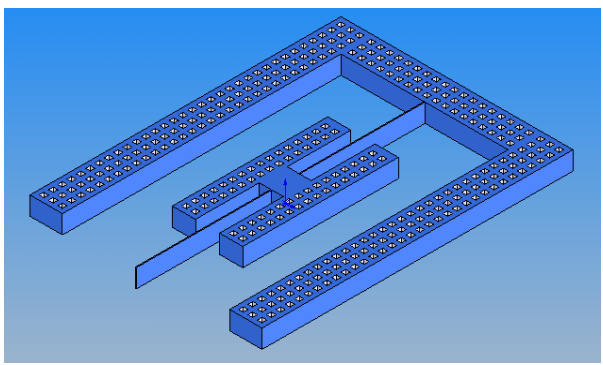

*Fig. 7.2.1: The view of the simulated structure with holes*

Simulations show that mode of resonance close to the required flexural resonant mode (Fig.7.2.2) was achieved typically at a frequency of 12590Hz at 25°. This mode was at the first rank, whose frequency was far from higher order modes: more than a few kHz in difference with the second mode. The resonance occurs with a slight indent in the y-axis, which was observed during simulation. The results shown in (Fig.7.2.2) summarize the simulation results for the temperature dependence of the first flexural resonance frequency. The relative variation of frequency versus temperature has either a positive or negative slope depending on the value of beam length.

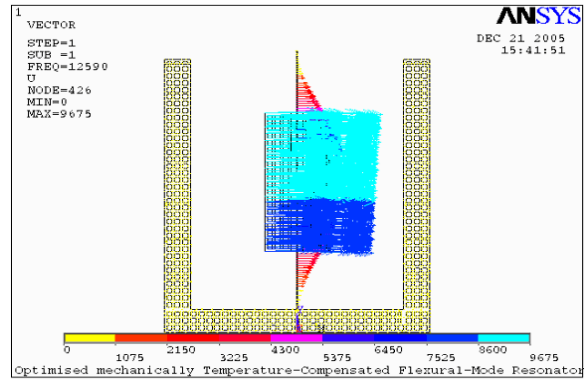

*Fig. 7.2.2: The simulated results to show the temperature stability of the structure*

## 7.3. Temperature Stability

The third level of simulation focused to achieve temperature stability closer in dimension to the fabricated structure. An optimisation process overlapped with the post process simulation was felt in the results of the simulation. This leads to the optimisation of the structure for the required resonant frequency with improved temperature stability.

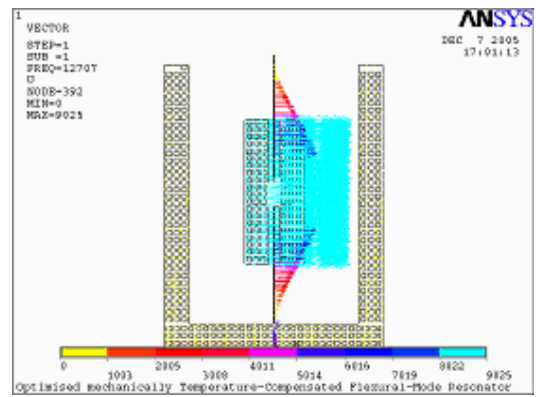

*Fig. 7.3.1: The simulated results to show the temperature stability of the structure.*

With respect to the previous structure there was a slight optimisation in the dimension of the structure. Though there was a very small change in the dimension, there was a significant effect on the temperature stability. The simulations have shown that the required flexural resonant mode (Fig.7.3.1) was achieved typically at a frequency of 12707 Hz. This mode was at the first rank, whose frequency was far from higher order modes: more than a few kHz in difference with the second mode. There was no relative variation of frequency versus temperature. One can also see that with the temperature stability was obtained for the ratio L1/L2 = 540/560 (=0.964),

## 8. CONCLUSION

This work has shown preliminary results on SOI electrostatically driven MEMS resonators with improved temperature stability. ANSYS simulations have shown that a frequency temperature coefficient in the order 1 ppm/°C is attainable based on the assumption of precision of 10 microns in structure length. First experimental results have shown a measured coefficient of 29.3 ppm/°C on non-optimised structures. Characterization of optimised structures is underway. Prospective work is the study of the impact of crystal orientation in the temperature behavior.